\documentclass[sigconf]{acmart}
\usepackage{subfig}
\usepackage{url}
\AtBeginDocument{%
  \providecommand\BibTeX{{%
    \normalfont B\kern-0.5em{\scshape i\kern-0.25em b}\kern-0.8em\TeX}}}

\setcopyright{none}

\renewcommand\footnotetextcopyrightpermission[1]{} 

\begin{document}

\fancyhead{}

\title{Hidden Author Bias in Book Recommendation}
\titlenote{Copyright 2022 for this paper by its authors. Use permitted under Creative Commons License Attribution 4.0 International (CC BY 4.0).\newline Accepted at FAccTRec 2022.}
\author{Savvina Daniil}
\email{s.daniil@cwi.nl}
\affiliation{%
  \institution{Centrum Wiskunde \& Informatica}
  \country{Amsterdam, The Netherlands}
}

\author{Mirjam Cuper}
\email{mirjam.cuper@kb.nl}
\affiliation{%
  \institution{National Library of the Netherlands}
  \country{The Hague, The Netherlands}
}
\author{Cynthia C.~S. Liem}
\email{c.c.s.liem@tudelft.nl}
\affiliation{%
  \institution{Delft University of Technology}
  \country{Delft, The Netherlands}
}
\author{Jacco van Ossenbruggen}
\email{jacco.van.ossenbruggen@cwi.nl}
\affiliation{%
  \institution{Vrije Universiteit Amsterdam}
  \country{Amsterdam, The Netherlands}
}
\author{Laura Hollink}
\email{l.hollink@cwi.nl}
\affiliation{%
  \institution{Centrum Wiskunde \& Informatica}
  \country{Amsterdam, The Netherlands}
}

\renewcommand{\shortauthors}{Daniil}

\begin{abstract}

Collaborative filtering algorithms have the advantage of not requiring sensitive user or item information to provide recommendations. However, they still suffer from fairness related issues, like popularity bias. In this work, we argue that popularity bias often leads to other biases that are not obvious when additional user or item information is not provided to the researcher. We examine our hypothesis in the book recommendation case on a commonly used dataset with book ratings. We enrich it with author information using publicly available external sources. We find that popular books are mainly written by US citizens in the dataset, and that these books tend to be recommended disproportionally by popular collaborative filtering algorithms compared to the users' profiles. We conclude that the societal implications of popularity bias should be further examined by the scholar community.  

\end{abstract}



\maketitle

\section{Introduction}
In recent years, Trustworthy AI is discussed as an important principle within the AI scholar community, but also on a national and international governmental level \cite{floridi2019establishing}. Robust algorithms have proven to be vulnerable to undesirable data patterns, and developers are not always able to beforehand recognize that. Public organizations specifically bear immense responsibility when utilizing AI to promote and facilitate their purposes. Recently publicized scandals have showcased that issues of bias and fairness are not always properly taken into account \cite{henley_booth_2020, biddle_saleh_2021}.

Recommender Systems are a very popular class of algorithms within domains like e-commerce and entertainment platforms \cite{linden2003amazon}. They often process an enormous amount of data in order to profile users and suggest products that they are likely to consume, and have proven to be very efficient. Collaborative filtering algorithms are a subset of Recommender Systems that specifically function by receiving consumption history as input rather than sensitive personal information of the users or items \cite{koren2022advances}. One might think that by omitting sensitive information, unwanted bias has no way to manifest in such a system. This is not the case; collaborative filtering approaches are still known to suffer from popularity bias. In short, popularity bias is an algorithmic effect where items that were originally popular in the training dataset tend to be recommended more often and thus have their popularity increase further. 

Popularity bias is recognized in both social and academic circles as a problem to be dealt with. It leads to decreased exposure of long-tail items that are critical for providers and platforms \cite{10.1145/3306618.3314309}, and users with niche tastes may have their interests ignored \cite{moviepaper}. At the same time, feedback loops are a point of concern for recommendations as homogeneity in user behavior amplified by popularity bias hinders utility \cite{chaney2018algorithmic}.

However, certain indirect implications of it are not placed in the forefront of said discussions. Given the fact that popular items tend to become more popular, it is appropriate to wonder which items were popular in the first place. We argue that popularity is often linked to other properties of an item, which translates to popularity bias leading to other sorts of data bias as well.

Specifically, we are investigating the topic of bias within the book recommendation domain. In this context, book authors are potential recipients of bias by a recommender system, as their popularity might coincide with sensitive characteristics such as age, ethnicity, religion, gender identity, sexual orientation, nationality and political preference. Multiple of said characteristics are known to instigate unfair treatment of authors in the publishing world \cite{baker_2020}. The treatment of authors by a recommender system based on their gender identity has received some attention from the scholar community \cite{ekstrand2021exploring, saxena2021exploring}, but other characteristics remain unexamined. 

In this specific work, we are zooming in on author country of citizenship. In a globalized book market, readers are no longer only exposed to local production. Steiner \cite{steiner2018global} argues that "..shifts in Internet use, the expansion of an e-book market, and the influence of a few large American corporations together appear to be transforming international publishing and retail". Hence, it is relevant to examine the relation between author country of citizenship and popularity, as well as the resulting effects in recommendation. Specifically, we are answering the following research questions:

 \begin{enumerate}
    \item Do commonly used recommender algorithms propagate data bias towards author country of citizenship? \label{rq1}
    \item What is the relation between author country of citizen bias and popularity bias? \label{rq2}
\end{enumerate}

We are considering these questions in the context of a dataset  with book ratings frequently used in literature to evaluate recommender systems.


\section{Related Work}
\subsection{Popularity bias}

Popularity bias is a long standing problem within the context of Recommender Systems and Information Retrieval \cite{brynjolfsson2006niches}. As explained by \cite{steck2011item} it stems from the users' tendency to provide feedback to popular items more often than long-tail items, which may not represent their true preferences. A system that recommends solely popular items to every user might score high on accuracy metrics, but it does not satisfy other requirements like item coverage \cite{mcnee2006being}. 

Recently, studies in the movie, music and book domain showed that popularity bias impacts different user groups disproportionally depending on their affinity for popular items \cite{moviepaper, bookpaper, musicpaper}. Specifically, users who tend to prefer niche items still receive mostly popular items as recommendations by well-known recommender algorithms. In that sense, niche users are treated unfairly by the system that fails to expose them to items that presumably would match their preferences better.

Popularity bias is recognized as a flaw of the system that leads to overall poor functionality in terms of fairness and balance, which transcend the basic goal of user satisfaction. Abdollahpouri argues in \cite{abdollahpouri2020popularity} that recommenders exist in multi-stakeholder platforms, thus rendering the satisfaction of interests, economic or otherwise, of item providers and platforms equally desirable. However, to our knowledge the societal connotations of popularity bias are not sufficiently addressed. Commonly used datasets with ratings tend to not contain information that would be needed to examine which items are consistently popular or which social groups of users/providers are unfairly treated. In other words, the potential of societal harm when popularity correlates with sensitive data properties adds an extra layer to the view of popularity bias breaching fairness.

\subsection{Bias in the book market}

The existence of different forms of bias in the publishing industry is well known. Female authors at large have historically been undervalued in the publishing landscape, for example by being less likely to be considered for reviews or major awards \cite{finn2016pseudonymous}. The 2020 academic study "Rethinking Diversity in Publishing" conducted qualitative interviews with professionals from all major UK literary agencies and concluded that publishers cater mostly to white middle-class audience and hence are less likely to invest in acquisition and promotion of authors of color \cite{saha_van_lente_2020}. Similar patterns have been reported in the US by various surveys \cite{author_demographics, leeandlowbooks_2022}. 

In the globalized book market, the US holds the largest revenue share according to a 2018 survey by the International Publishers Association \cite{ipareport}. At the same time, 7 out of the 10 highest-paid authors worldwide reported in 2018 by Forbes are US citizens \cite{cuccinello_2018}. Social cataloging websites commonly used by younger readers for discovering and reviewing new titles also tend to be US-centric. For example, the American Amazon owened Goodreads places first in the list of websites related to "Libraries and Museums" ranked by traffic \cite{similarweb}. While US based Goodreads users make up an estimated 40\% of the traffic \cite{walsh2021goodreads}, the popular authors' country of origin is disproportionally skewed, with 12 out of the 15 most often rated authors between 2016 and 2021 being American \cite{martin_2021}. In other words, the US holds strong influence in the modern literary world, particularly when it comes to renowned authors.

\section{Research Design}

\subsection{Data}

In order to study hidden author bias in book recommendation, we selected the Book-Crossing dataset \cite{bookcrossing}, a very popular dataset for book recommendations created in 2004 by crawling the website of Book-Crossing for four weeks. Book-Crossing is an international book exchange community whose members can leave books anywhere in the world and indicate their location to other users. The users can then update their profile with the information that they read a new book and also submit a rating from 1 to 10. 

\begin{figure*}[!htbp]
  \centering
  \includegraphics[width=\textwidth]{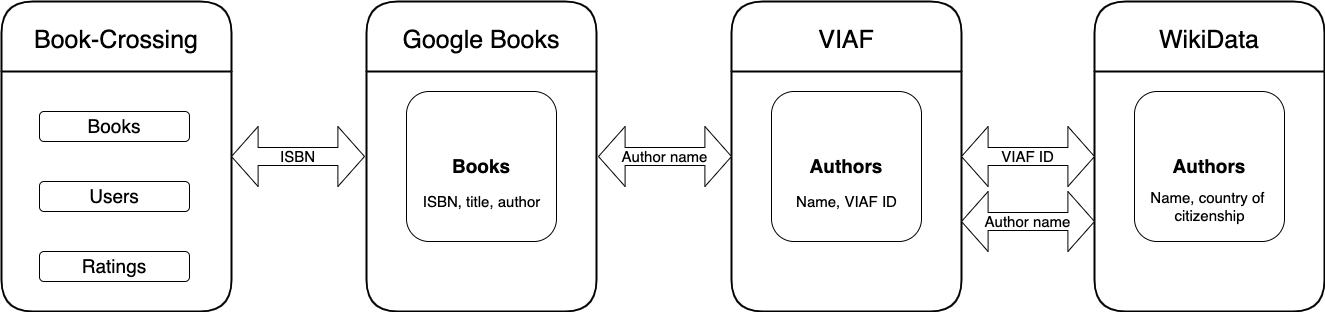}
    \caption{Links between the datasets.}
  \label{fig:data}
\end{figure*}

The dataset consists of three sub-datasets: book ratings, users and their location, and items along with some information on them, namely the name of the author, the publisher, and the year of publication. We hence used external sources to enrich it with additional author information when publicly available. We took the following steps, as shown in Figure \ref{fig:data}:
\begin{enumerate}
    \item We linked Book-Crossing to Google Books based on ISBN. We used the author name included in Google Books to validate and/or correct the one provided in Book-Crossing, since we noticed mistakes and inconsistencies.
    \item We linked to Virtual International Authority File (VIAF) using author name. 
    \item We linked to WikiData using VIAF ID, and validated using author name. From WikiData, we extracted author country of citizenship.
\end{enumerate}

Naghiaei et al. \cite{bookpaper} utilized the Book-Crossing dataset in their Fairbook system to investigate the unfairness of popularity bias in the book recommendation case. In order to build on their work by comparing how certain recommender algorithms propagate popularity versus how they propagate author bias, we reproduced their preprocessing approach:
\begin{itemize}
    \item Remove implicit ratings.
    \item Remove users with more than 200 ratings.
    \item Remove users with less than 5 ratings.
    \item Remove items with less than 5 ratings.
\end{itemize}

The application of such cut offs is common in the recommendation literature given the vast but sparse amount of data that online platforms usually possess \cite{lu2012recommender}. Therefore, it is compelling to observe whether it has an effect on the inherent bias of the dataset.  

We further processed the dataset to account for duplicate items with different ISBN numbers, as well as ISBN numbers included in the ratings dataset but not in the items dataset that we considered to be mistakes. This processing is documented and accessible in our GitHub. The general characteristics of the processed version of the entire dataset, as well as of the Fairbook dataset can be seen at Table \ref{tab:dataset}. 

\begin{table}[!htbp]{Dataset characteristics}
\begin{tabular}{l|l|l|l|}
\cline{2-4}
\multicolumn{1}{c|}{}                                       & \textbf{\#ratings} & \textbf{\#books} & \textbf{\#users} \\ \hline
\multicolumn{1}{|l|}{\textbf{Entire dataset - processed}}   & 1,021,847          & 222,496          & 92,106           \\ \hline
\multicolumn{1}{|l|}{\textbf{Fairbook dataset - processed}} & 86,356             & 5,504            & 6,354            \\ \hline
\end{tabular}
\caption{The cut off thresholds introduced by Fairbook significantly decrease the dataset size.}
\label{tab:dataset}
\end{table}

\subsection{Measuring bias}\label{subsec:measuring-bias}
In the interest of answering the research questions, we devised a way to measure the country of citizenship bias existing in the data and compare it to the recommendations offered when training different recommender algorithms on that data. We applied the research design to the Fairbook dataset in order to juxtapose the propagation of data bias to the propagation of popularity bias studied by the Fairbook system.

First, we examined whether there is a significant relationship between book popularity and country of citizenship of the author. We define popularity of an item to be the amount of ratings of said item, similarly to Abdollahpouri et al. \cite{moviepaper}. Afterwards, we trained the same algorithms as the Fairbook system so we can directly compare the results. Specifically, we trained 11 recommender algorithms, 9 collaborative filtering and two dummy approaches as seen in Table \ref{Algorithms}, with a 80-20\% split. We then recommended to each user 10 books as a result of each algorithm's predicted ranking.

\begin{table}[!htbp]
\centering
\resizebox{\columnwidth}{!}{\begin{tabular}{ccc}
\hline
\multicolumn{1}{|c|}{\textbf{Algorithm}}                     & \multicolumn{1}{c|}{\textbf{Approach}} & \multicolumn{1}{c|}{\textbf{Acronym}} \\ \hline
\textbf{User K-Nearest Neighbors}                            & K-Nearest Neighbors                    & \textbf{UserKNN}                      \\
\textbf{Matrix Factorization}                                & Matrix Factorization                   & \textbf{MF}                           \\
\textbf{Probabilistic Matrix Factorization}                  & Matrix Factorization                   & \textbf{PMF}                          \\
\textbf{Non-negative Matrix Factorization}                   & Matrix Factorization                   & \textbf{NMF}                          \\
\textbf{Weighted Matrix Factorization}                       & Matrix Factorization                   & \textbf{WMF}                          \\
\textbf{Hierarchical Poisson Factorization}                  & Matrix Factorization                   & \textbf{PF}                           \\
\textbf{Bayesian Personalized Ranking  }                              & Ranking-based                          & \textbf{BPR}                          \\
\textbf{Neural Matrix Factorization}                         & Neural Network-based                   & \textbf{NeuMF}                        \\
\textbf{Variational Autoencoder for Collaborative Filtering} & Neural Network-based                   & \textbf{VAECF}                        \\
\textbf{Most Popular}                                        & Dummy                                  & \textbf{MostPop}                      \\
\textbf{Random}                                              & Dummy                                  & \textbf{Random}                      
\end{tabular}}
\caption{Recommender algorithms chosen to be trained on our dataset. They are the same ones trained by the Fairbook system.}
\label{Algorithms}
\end{table}

Finally, we compared the distribution of author country of citizenship in the users' profiles versus recommended lists for every algorithm. That way we were able to observe whether and how each algorithm's recommendations impact this distribution and compare the result to the propagation of popularity bias for each algorithm.

Our code both for the data enrichment\footnote{\url{https://github.com/SavvinaDaniil/EnrichBookCrossing}} and the bias analysis\footnote{\url{https://github.com/SavvinaDaniil/BiasInRecommendation}} have been made open source.

\section{Results \& Discussion}
\subsection{Bias in the data}

\begin{figure}[!htbp]
  \centering
  \subfloat[Country distribution in the entire book dataset.]{\includegraphics[scale=0.17]{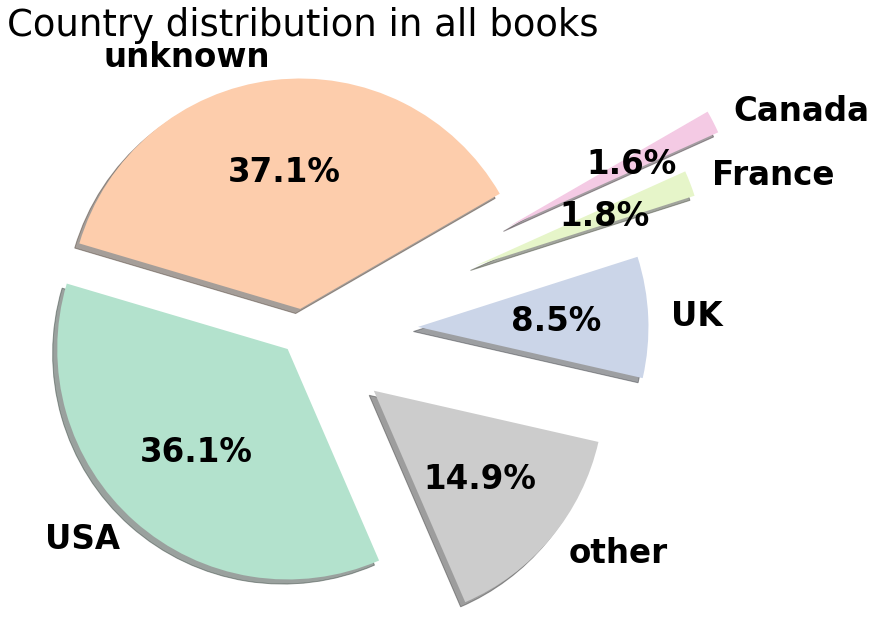}\label{fig:books1}}
  \hfill
  \subfloat[Country distribution in the book dataset with Fairbook cut offs.]{\includegraphics[scale=0.17]{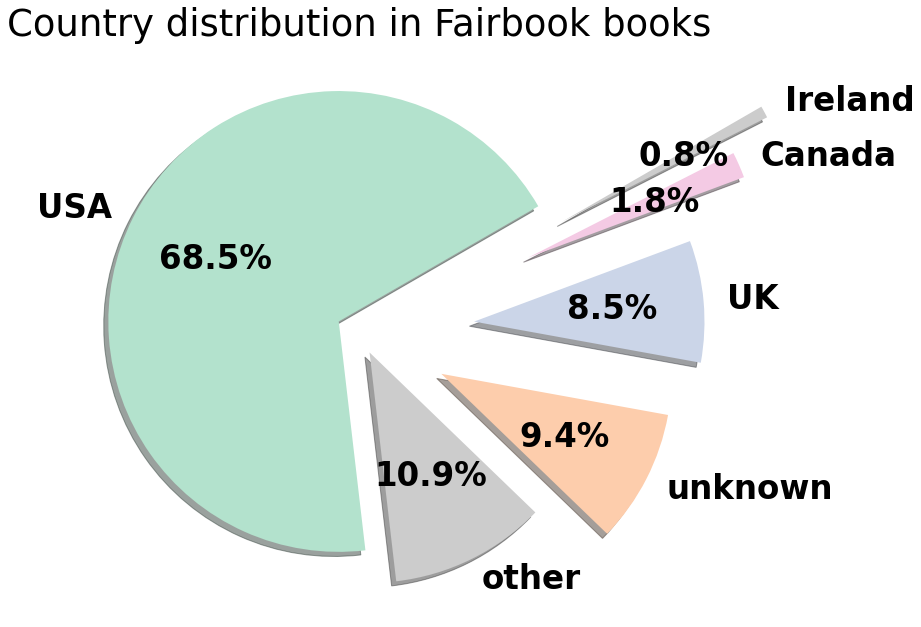}\label{fig:books2}}
  \caption{Country distribution in unique books.}
\end{figure}

\begin{figure}[!htbp]
  \centering
  \subfloat[Country distribution in the entire ratings dataset.]{\includegraphics[scale=0.17]{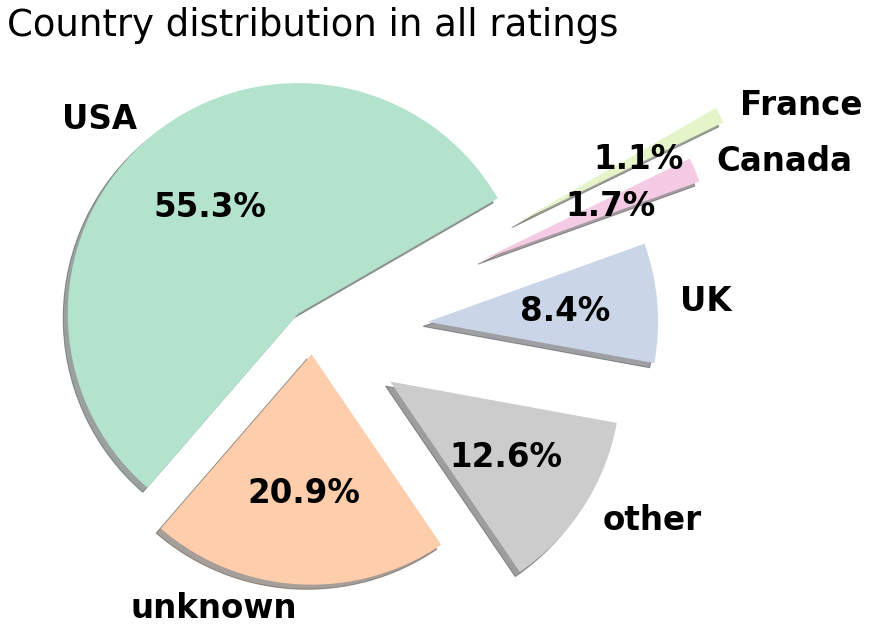}\label{fig:ratings1}}
  \hfill
  \subfloat[Country distribution in the ratings dataset with Fairbook cut offs.]{\includegraphics[scale=0.17]{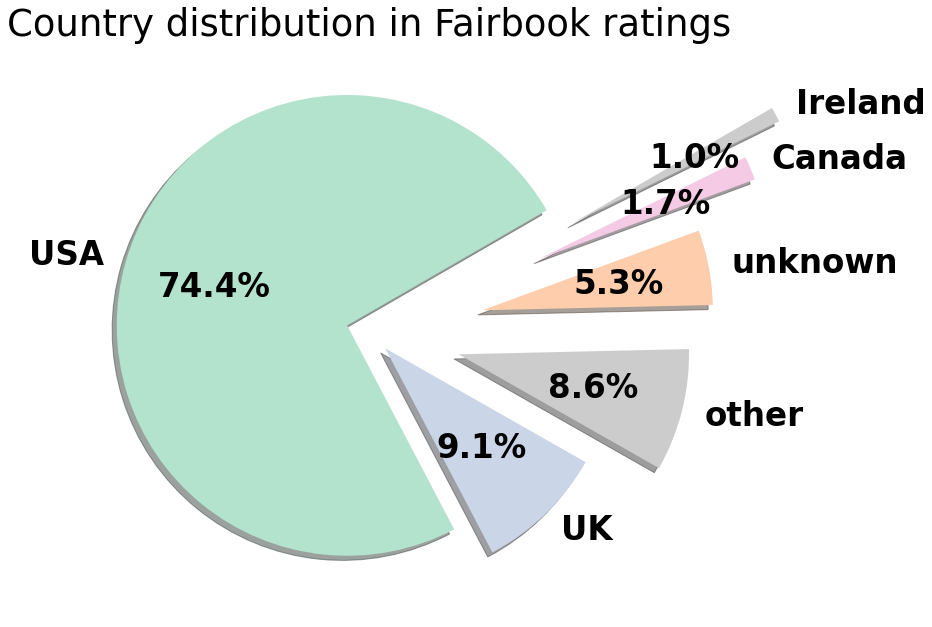}\label{fig:ratings2}}
  \caption{Country distribution in unique ratings.}
\end{figure}

Data analysis shows clear bias in favor of US citizens authors. As seen in figure \ref{fig:books1}, in the entire set of unique books around 36\% were written by American authors. When the cut offs are introduced, the percentage of American-authored books grows to 68.5\%, as seen in figure \ref{fig:books2}. The same effect appears in the ratings datasets, figures \ref{fig:ratings1} and \ref{fig:ratings2}. Out of the entire dataset, 55.3\% of the ratings are given to books written by US citizens. Similarly, the phenomenon is more apparent for the Fairbook ratings, with the percentage growing to 74.4\%.

We conclude that introducing the cut offs increased the bias of the dataset in favor of US citizens authors. This remark already implies a direct relation between item popularity and country of citizenship of the author, given that Fairbook excludes all items with less than 5 raters. At the same time, the Book-Crossing website reports that 20\% of the users are based in the US \cite{book-crossing}. We can deduce that American authors are overrepresented in the data compared to the amount of American users.

Finally, we divided the books on American-authored and non American-authored. A t-test between the mean popularity of the two sets showed a significant higher popularity for the American-authored items. 

\subsection{Bias in recommendation}
\begin{figure}[!htbp]
  \centering
  \includegraphics[width=0.5\textwidth]{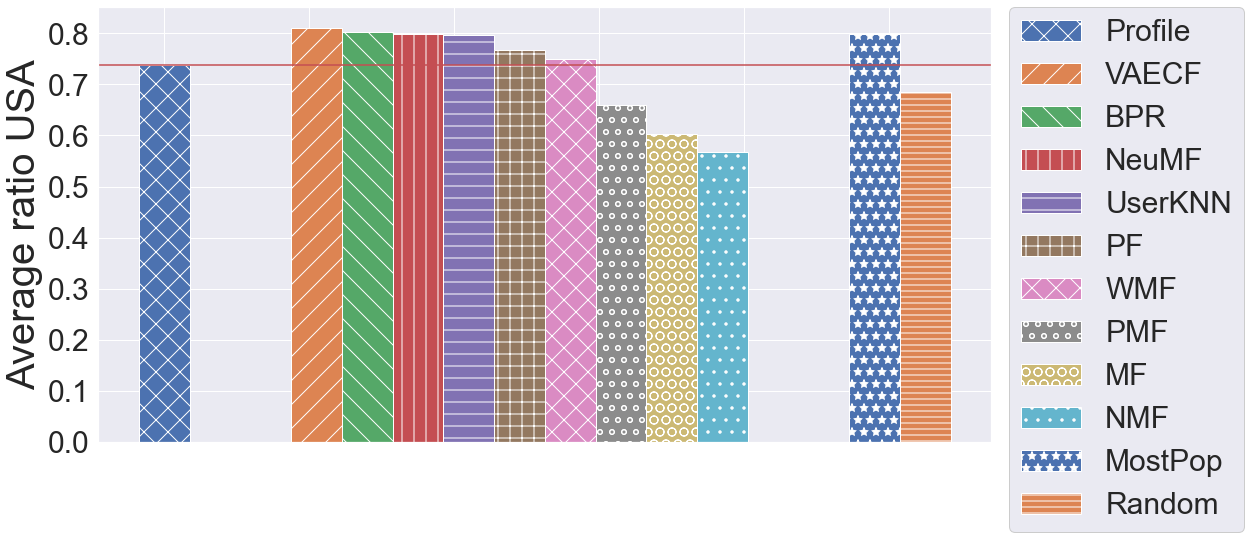}
    \caption{Average ratio of recommended books by every algorithm that were written by US citizens. Comparison with the average ratio of American-authored books in the users' profiles. }
  \label{fig:results_10}
\end{figure}

After applying the recommendation process as explained in \ref{subsec:measuring-bias}, we compared the ratio of American-authored books in users' profiles to the one in the recommended list of each algorithm. Figure \ref{fig:results_10} shows that, setting aside Random and MostPopular, all collaborative filtering recommenders increase the ratio, except from NMF, MF, and PMF.

\begin{figure}[!htbp]
  \centering
  \includegraphics[width=0.5\textwidth]{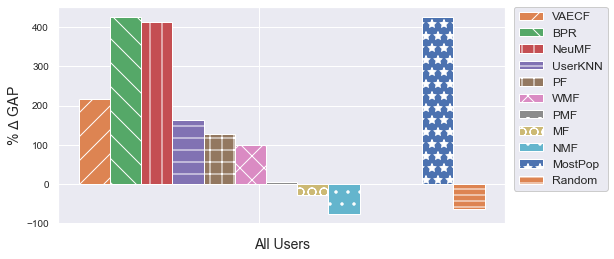}
    \caption{Relative increase in average popularity between profile and recommendation by every algorithm, averaged over all users.}
  \label{fig:results_fairbook}
\end{figure}

In order to compare these algorithmic tendencies to popularity bias, we used the \%$\Delta$GAP metric \cite{moviepaper}. It expresses the relative increase of average item popularity in a user's recommended list compared to their profile. Figure \ref{fig:results_fairbook} is a recreation of a graph from Fairbook, with the difference that \%$\Delta$GAP is calculated over all users rather than per user group. It shows that all collaborative filtering algorithms but PMF, MF, and NMF produce recommended lists with increased average item popularity. In other words, the same algorithms that recommend excessively American-authored books also recommend excessively popular books.

\begin{figure*}[!htbp]
  \centering
  \subfloat[Random]{\includegraphics[width=0.3\textwidth]{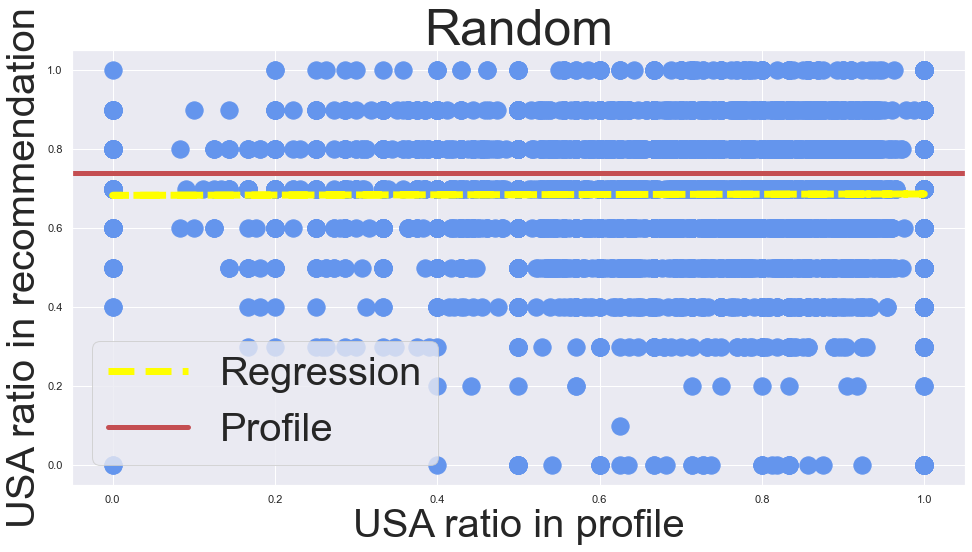}\label{fig:random}}
  \subfloat[MostPop]{\includegraphics[width=0.3\textwidth]{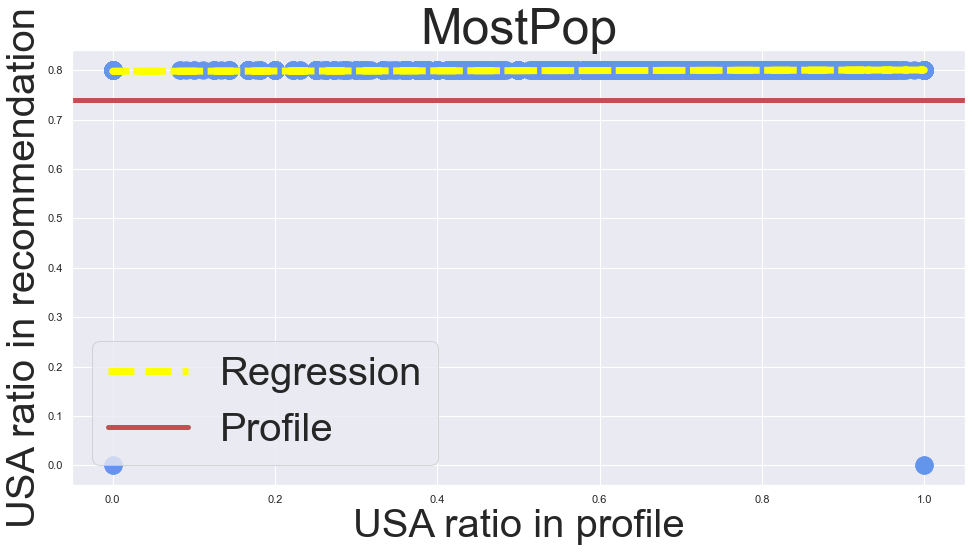}\label{fig:mostpop}}
  \subfloat[UserKNN]{\includegraphics[width=0.3\textwidth]{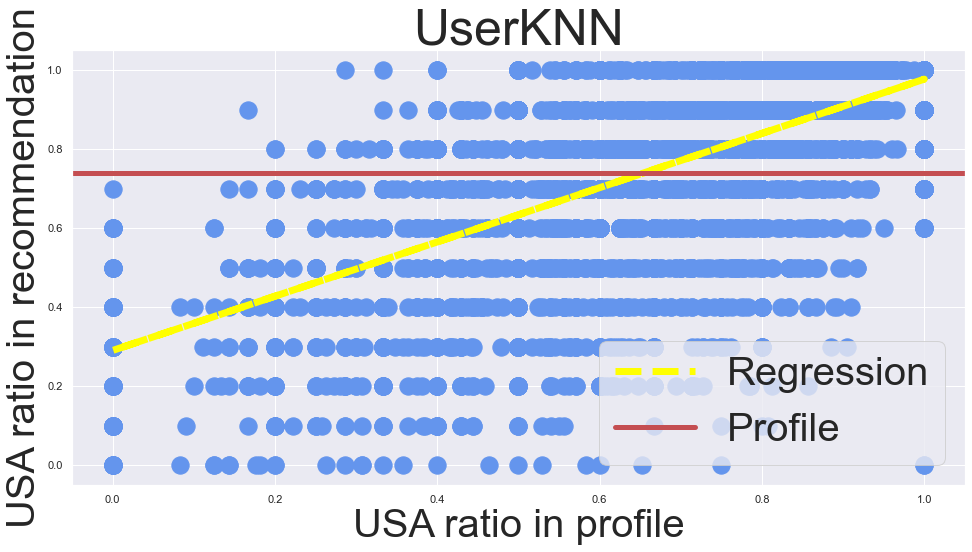}\label{fig:userknn}}
  \hfill
  \subfloat[MF]{\includegraphics[width=0.3\textwidth]{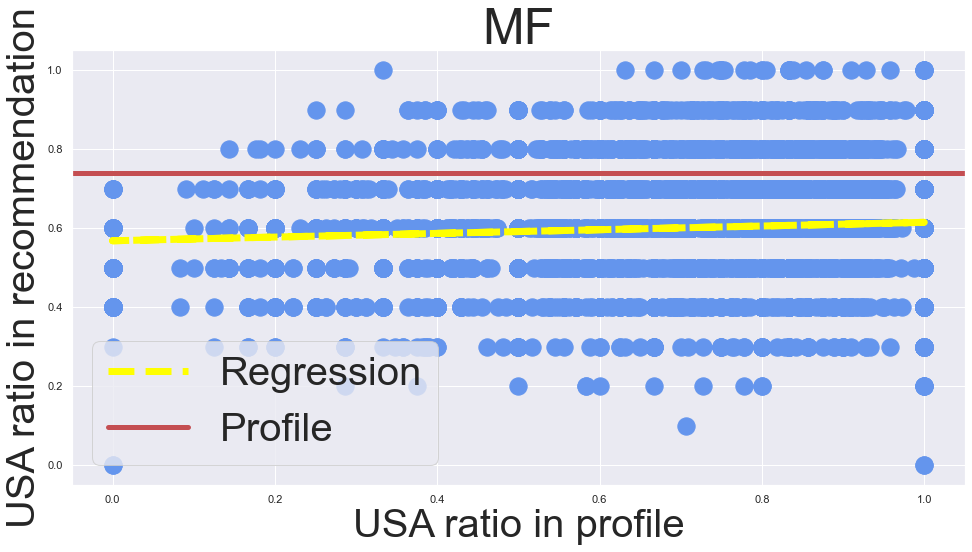}\label{fig:mf}}
  \subfloat[PMF]{\includegraphics[width=0.3\textwidth]{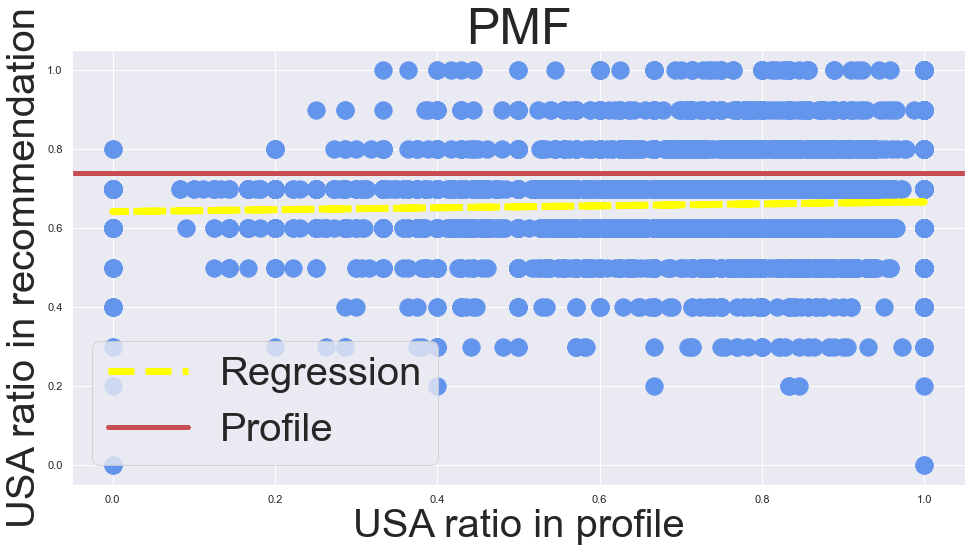}\label{fig:pmf}}
  \subfloat[NMF]{\includegraphics[width=0.3\textwidth]{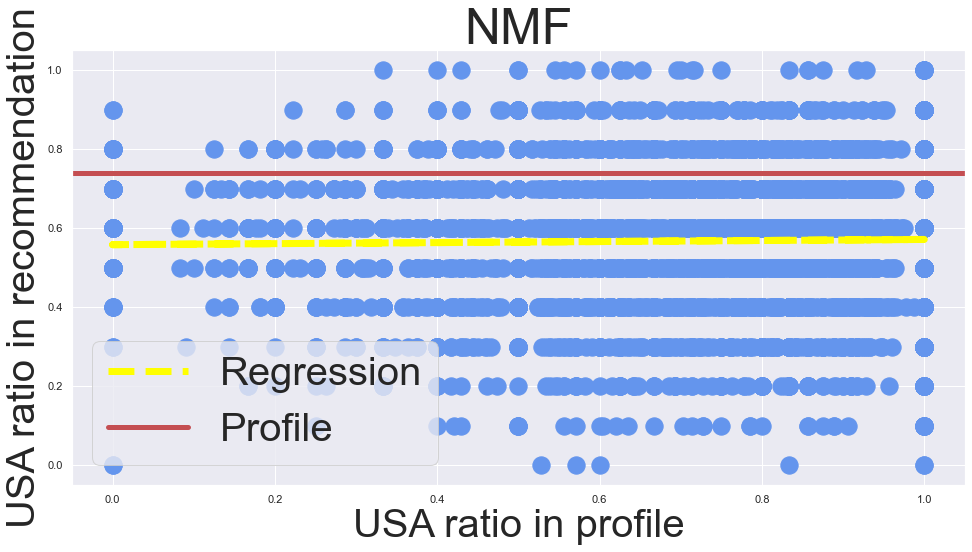}\label{fig:nmf}}
  \hfill
  \subfloat[WMF]{\includegraphics[width=0.3\textwidth]{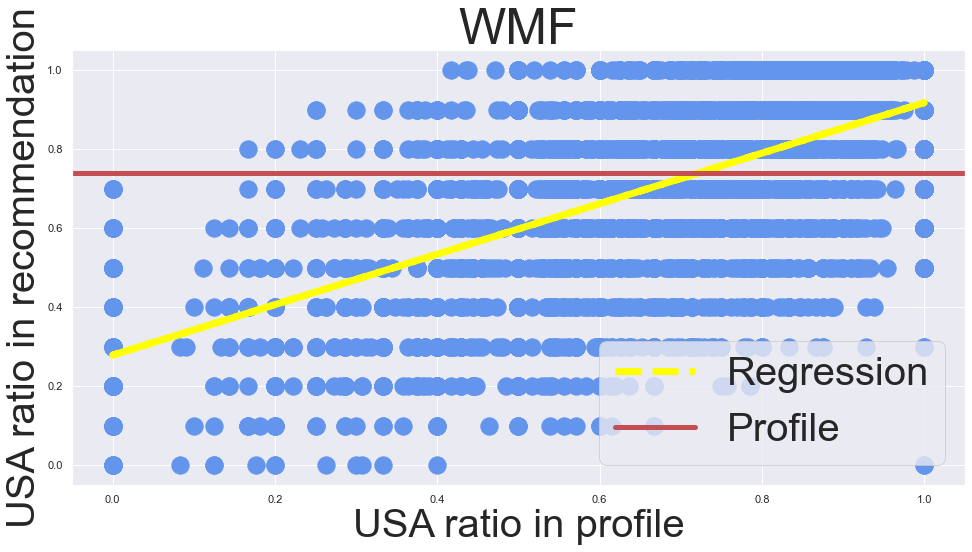}\label{fig:wmf}}
  \subfloat[BPR]{\includegraphics[width=0.3\textwidth]{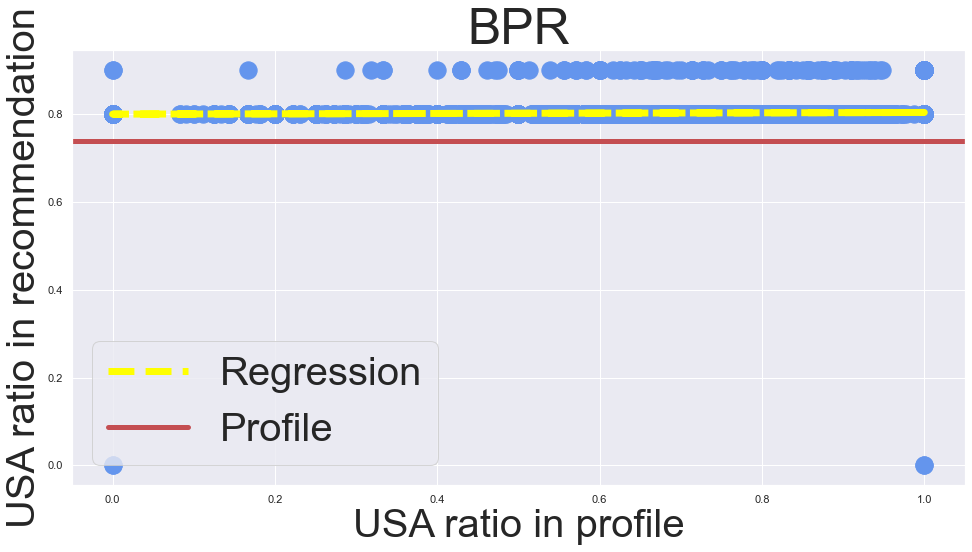}\label{fig:bpr}}
  \subfloat[PF]{\includegraphics[width=0.3\textwidth]{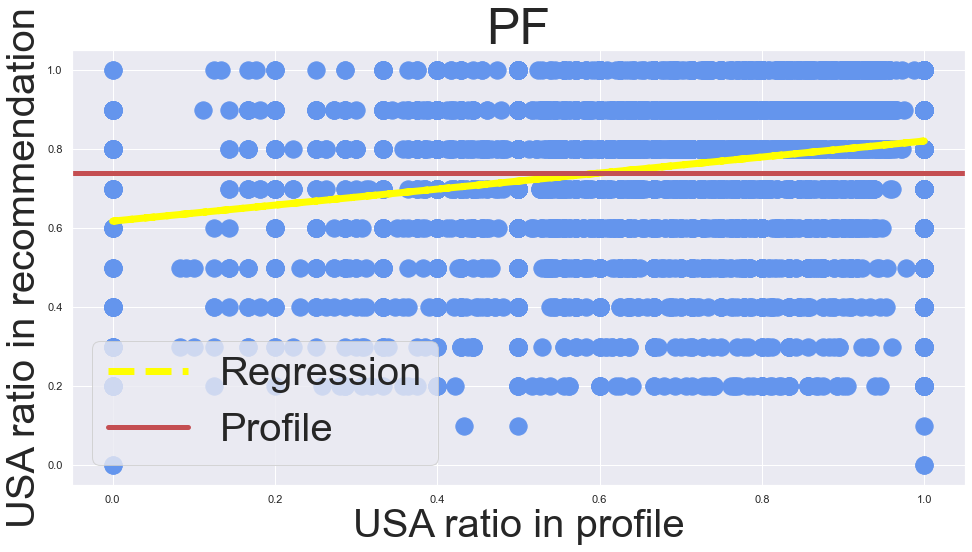}\label{fig:pf}}
\hfill
  \subfloat[NeuMF]{\includegraphics[width=0.3\textwidth]{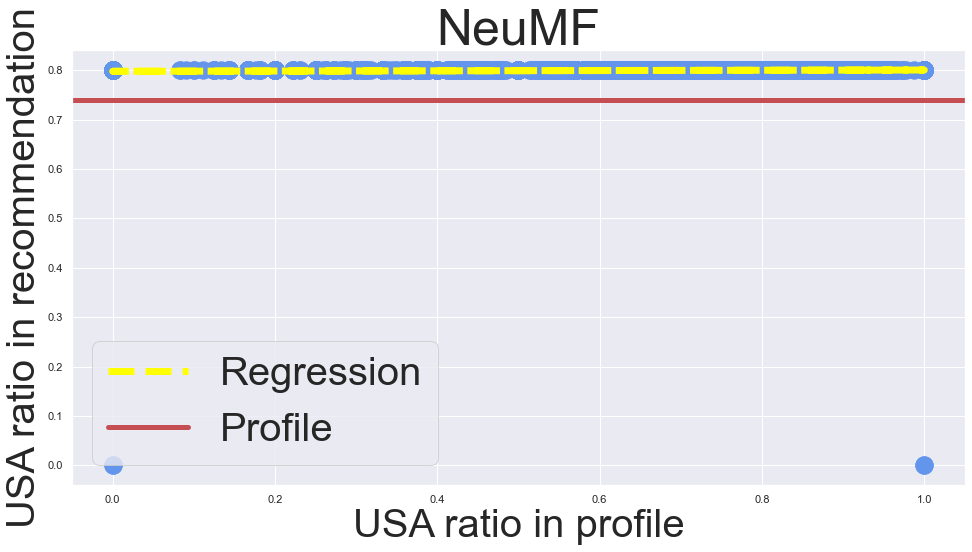}\label{fig:neumf}}
  \subfloat[VAECF]{\includegraphics[width=0.3\textwidth]{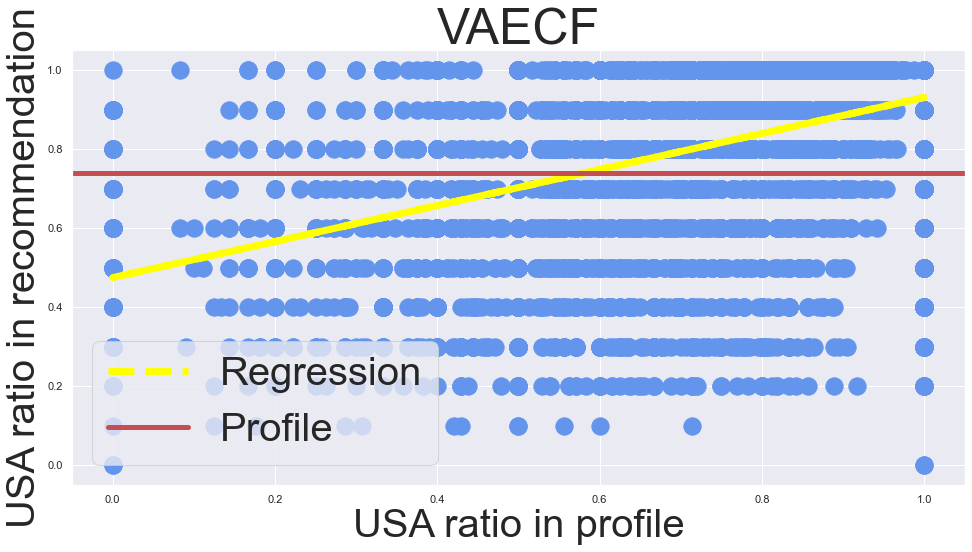}\label{fig:vaecf}}
  
  \caption{Ratio of American-authored books in profile versus in recommendation for every user.}
  \label{fig:everyuser}
\end{figure*}

We then zoom in on each user and compare their profile ratio of US citizens authors to the recommended list of each recommender algorithm. Figure \ref{fig:everyuser} shows that most collaborative filtering algorithms either recommend consistently disproportionally many American-authored books despite the user's preferences (BPR, NeuMF), or they show a correlation between profile and recommended ratio (UserKNN, WMF, PF, VAECF). The only exceptions are NMF, MF, and PMF, which consistently recommend lower ratio than average. UserKNN, WMF, PF, and VAECF may show correlation between ratio of American-authored books in user profile and recommendation, but as seen in Figure \ref{fig:results_10} they on average increase the ratio.

When comparing our results with the Fairbook conclusions, we see that they follow similar trends in the sense that the same recommender algorithms that propagated popularity bias also acted in favor of American-authored items by disproportionally recommended them. Naghiaei et al. stated that "...no positive correlation exists in PMF, MF, and NMF, indicating that the latter algorithms in Matrix Factorization-based approaches are not prone to popularity bias in Book-Crossing dataset.". This is a strong indication that the behavior we observe in our results is a direct result of popularity bias.

\subsection{Analysis}
Given our results, we can answer Research Question \ref{rq1} by stating that certain recommender algorithms propagate bias against author country of citizenship, despite not receiving country of citizenship as a feature in the training process. Specifically, American authors were favoured by those algorithms by being disproportionally recommended. Comparing propagation of American author bias and popularity bias also enlightens us on its origins. Considering the similar manifestations of bias by the different recommenders, we do believe that the observed author bias was directly incited by the known phenomenon of popularity bias, thus answering Research Question \ref{rq2}. 

The results indicate that American book platforms accessed by users around the world may be disproportionally US centric, and that this phenomenon can get perpetuated when certain recommenders are being applied without proper examination. Lack of geodiversity in open data is a known problem that requires addressing when this data is employed for either research or commercial purposes \cite{shankar2017no}. The results also hint that popularity bias in recommendation should be viewed as more than an issue of potentially weak performance that mostly impacts e-commerce. Naturally, popularity is not always a bad criterion for recommendation, as argued by \cite{zhao2021popularity}. Word-of-mouth is commonly how we as humans stay informed on art, politics, science, and other socially important topics. Popularity among topics can signal what is relevant and/or of quality, and automated systems can often accurately encode what is typically happening in society. 

That being said, AI is known to track patterns in the data that do not objectively depict reality, but rather often undesirable historical context. Consequently, employing AI in a way to directly imitate the data without check points can bring about societal harm in unexpected ways. In the case of book recommendation, efficient recommender algorithms can be used to increase reading habits, but without risking demoting certain books by virtue of their authors' sensitive features that may correlate with past unpopularity. For this reason, our work indicates that researchers must be very thorough when examining potential bias in recommended systems and make active effort to address it, even when seemingly popularity is the cause.

\section{Conclusion \& Future Work}
This paper is a part of an ongoing attempt to reflect on the concepts of bias and inclusivity in the book recommendation case. We examined the phenomenon of hidden bias in recommendation introduced by commonly used recommender algorithms that do not take any item features as input. We theorized that feature bias comes as a direct result of popularity bias which is a known issue that recommenders face. We investigated the hypothesis in the context of book recommendations and used a well-known book ratings dataset to validate our hypothesis. We found the books written by American authors within the dataset to be significantly more popular compared to the rest. We also found that many commonly used collaborative filtering algorithms on average recommend more American-authored books than in the users' profile. In fact, the same algorithms seem to propagate popularity bias according to previours work on the same dataset.

We believe that hidden bias in recommender systems should be in the spotlight for the scholar community. Fair treatment of social groups should be a priority for AI developers. As shown, it is not sufficient to explicitly exclude sensitive personal features from the training process; bias can appear and manifest in proxies. 

In collaboration with the National Library of the Netherlands, we wish to study recommender systems in terms of their capability to be inclusive. In their published AI principles, under the "Inclusive" section the National Library acknowledges AI's susceptibility to bias. At the same time, they equate maintaining inclusivity to knowing "... where and to what extent bias occurs [in the data], so that it can be eliminated or compensated." \cite{kbprinciples} Given a library's unique position as a public organization that aims to promote education and responsibility to treat all social groups fairly, it is crucial to identify potential blind spots that can cause bias against author social groups to manifest when a recommender system is in use.

Currently we only compare profile with recommended lists, since we aimed to focus on the relation between popularity bias and author bias. In the future, it will be interesting to use feedback loops in order to investigate how the observed phenomenon progresses through multiple iterations of recommendation and consumption. 

At the same time, there is room for expanding the data enrichment process. We currently depend on the information existing in WikiData on someone's country of citizenship, which is evidently incomplete. Moreover, it is documented that WikiData can be biased in terms of country of citizenship, with Europe and North America being often overrepresented \cite{shaik2021analyzing}. The aforementioned skewness in the available WikiData entries may be introducing additional bias to our data. We plan to address this topic thoroughly in future work.

The question of how bias surfaces in book recommendation is not negligible; it could be directed to authors, books, users. Our study focuses on authors and specifically country of citizenship, but each of these dimensions can be important depending on the context and thus should be given attention. For future work, we plan to consider the other dimensions of the problem as well. By having a well rounded understanding of hidden bias in book recommendation, we can move on to properly account and compensate for it. In this case, libraries can benefit from the automation recommender systems offer in attracting users, while ensuring that their value of inclusivity is being properly adhered to.

\begin{acks}
Funded by the National Library of the Netherlands.
\end{acks}

\bibliographystyle{ACM-Reference-Format}
\bibliography{main}

\end{document}